\documentclass[prl,aps,showpacs,twocolumn]{revtex4}
\usepackage{amsfonts}
\usepackage{amsmath}
\usepackage{amssymb}
\usepackage[dvips]{graphicx}

\begin{document}

\title{Sudden Birth Versus Sudden Death of Entanglement in Multipartite Systems}
\date{\today}
\author{ C. E. L\'{o}pez$^{1}$, G. Romero$^{1}$, F. Lastra$^{2}$, E. Solano$^{3,4}$ and J. C. Retamal$^{1}$}
\affiliation{$^{1}$Departamento de
F\'{\i}sica, Universidad de Santiago de Chile, USACH, Casilla 307
Correo 2 Santiago, Chile
\\
$^{2}$Facultad de F\'{\i}sica,
Pontificia Universidad Cat\'{o}lica de Chile, Casilla 306,
Santiago 22, Chile
\\
$^3$Physics Department, ASC, and CeNS,
Ludwig-Maximilians-Universit\"at, Theresienstrasse 37, 80333
Munich,
Germany
\\
$^4$Departamento de Qu\'{\i}mica F\'{\i}sica, Universidad del Pa\'{\i}s Vasco - Euskal Herriko Unibertsitatea, Apdo. 644, 48080 Bilbao, Spain}
\pacs{03.65.Yz, 03.65.Ud, 03.67.Mn}

\begin{abstract}
We study the entanglement dynamics of two cavities interacting
with independent reservoirs. Expectedly, as the
cavity entanglement is depleted, it is transferred to the
reservoir degrees of freedom. We find also that when the cavity
entanglement suddenly disappear, the reservoir entanglement
suddenly and necessarily appears. Surprisingly, we show that this
{\it entanglement sudden birth} can manifest before,
simultaneously, or even after {\it entanglement sudden death}.
Finally, we present an explanatory study of other entanglement
partitions and of higher dimensional systems.
\end{abstract}

\maketitle

Dynamical behavior of entanglement under the action of the
environment is a central issue in quantum information~\cite{Cubitt0305,Diosi03,Dodd04,andre07}. Recently, it has
been observed that two qubits affected by uncorrelated reservoirs
can experience disentanglement in a finite time despite coherence
is lost
asymptotically~\cite{Diosi03,Dodd04,andre07,Yu04,Marcelo06}. This
phenomenon, called entanglement sudden death (ESD), has recently deserved a
great attention~\cite{Jamroz06,Ficek06,terra07,esd,Derkacz06,cpsun07,lastra07}, and has been observed in the lab for entangled
photon pairs ~\cite{almeida07}, and atomic
ensembles~\cite{laurat07}.

To our knowledge ESD has been studied mainly in relation to bipartite systems, while a deeper understanding is associated to the question of where does the lost entanglement finally go. This question would be properly
answered by enlarging the system to include reservoir degrees of
freedom. Intuitively, we may think that the lost
entanglement has to be transferred to the reservoir degrees of freedom. However, is this entanglement swapped continuously? If the bipartite entanglement suffers ESD, what
can we say about the transferred entanglement? Should there be a simultaneous {\it entanglement sudden birth} (ESB) on
reservoir states, or when would this entanglement be created?
In this work, we thoroughly study the entanglement transfer
from the bipartite system
to their independent reservoirs.  We show that ESD of a bipartite system state is intimately linked to ESB of entanglement between the reservoirs, though their apparition times follow counterintuitive rules.

To illustrate the problem we  have chosen the case of entangled
cavity photons being affected by dissipation, as in the case of two modes inside the same dissipative cavity or single modes in two different ones. The present study could certainly be extended to other physical systems like
matter qubits. First we study the case of qubits for two uncoupled (cavity)
modes having up to one photon. Then, we extend our
treatment to investigate wether or not the effect is present in
higher dimensions (qudits).

Since each mode evolves independently, we can learn how to characterize the
evolution of the overall system from the
mode-reservoir~dynamics. The interaction between a single
cavity mode and an $N$-mode reservoir is described through the
Hamiltonian
\begin{equation}
\hat{H}=\hbar \omega \hat{a}^{\dag }\hat{a}+\hbar
\sum_{k=1}^{N}\omega _{k} \hat{b}^{\dag }\hat{b}+\hbar
\sum_{k=1}^{N}g_{k}\left( \hat{a}\hat{b}_{k}^{\dag
}+\hat{b}_{k}\hat{a}^{\dag }\right) . \label{Hamiltonian}
\end{equation}
Let us consider the case when a cavity mode is containing a single
photon and its corresponding reservoir is in the vacuum state,
\begin{equation}
|\phi _{0}\rangle =|1\rangle _{\mathrm{c}}\otimes
|\bar{\mathbf{0}}\rangle _{\mathrm{r}},  \label{initial1}
\end{equation}
where, $|\bar{\mathbf{0}}\rangle
_{\mathrm{r}}=\prod_{k=1}^{N}|0_{k}\rangle _{\mathrm{r}}$. It is
not difficult to realize that the evolution given by~(\ref{Hamiltonian}) leads to the state
\begin{equation}
|\phi _{t}\rangle_{\mathrm{cr}} =\xi (t)|1\rangle
_{\mathrm{c}}|\bar{\mathbf{0}}\rangle
_{\mathrm{r}}+\sum_{k=1}^{N}\lambda _{k}(t)|0\rangle
_{\mathrm{c}}|1_{k}\rangle _{\mathrm{r}},  \label{psit1}
\end{equation}
where the state $|1_{k}\rangle _{\mathrm{r}}$ accounts for the
reservoir having one photon in mode $k$. The amplitude $\xi (t)$
converges to $\xi (t)=\exp{(-\kappa t / 2)}$ in the limit
of $N \rightarrow \infty$ for a reservoir with a flat spectrum. The
right-hand term of the last equation can be rewritten in terms of a
collective state of the reservoir modes as
\begin{equation}\label{phit1}
|\phi _{t}\rangle =\xi (t)|1\rangle
_{\mathrm{c}}|\bar{\mathbf{0}}\rangle _{\mathrm{r}}+\chi
(t)|0\rangle _{\mathrm{c}}|\mathbf{\bar{1}}\rangle _{\mathrm{r}} .
\end{equation}
Here, we defined the normalized collective state with one excitation
in the reservoir as
\begin{equation}
|\mathbf{\bar{1}}\rangle _{\mathrm{r}}=\frac{1}{\chi
(t)}\sum_{k=1}^{N} \lambda _{k}(t)|1_{k}\rangle _{\mathrm{r}},
\end{equation}
and the amplitude $\chi (t)$ in Eq. (\ref{phit1}) converge to the
expression $\chi (t)=({1-\exp{(-\kappa t)}}¡)^{1/2}$ in the large
$N$ limit. Described in this way the cavity and reservoir evolve as
an effective two-qubit system~\cite{Yonac07}.

\begin{figure}[t]
\includegraphics[width=55mm]{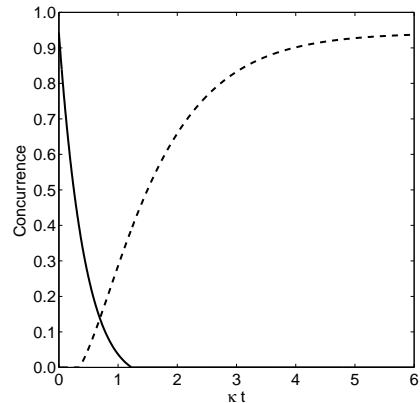}
\caption{Evolution of two-qubit concurrence
$\mathcal{C}_{\mathrm{c}_1 \mathrm{c}_2}$ (solid line) and
$\mathcal{C}_{\mathrm{r}_1 \mathrm{r}_2}$ (dashed line), for the
initial state of Eq.~(\ref{cavent}) with $\alpha=1/\sqrt 3$ and
$\beta=2/\sqrt 3$.}  \label{figqubit}
\end{figure}

Let us now study the joint evolution of two qubits with their
corresponding reservoirs initially in the global state
\begin{equation}\label{cavent}
|\Phi_0\rangle=(\alpha|0\rangle_{\mathrm{c}_1}
|0\rangle_{\mathrm{c}_2}+\beta|1\rangle_{\mathrm{c}_1}
|1\rangle_{\mathrm{c}_2})|\bar{\mathbf{0}}\rangle_{\mathrm{r}_1}
|\bar{\mathbf{0}}\rangle_{\mathrm{r}_2}.
\end{equation}
According to Eq.(\ref{phit1}), the evolution of the overall system
will be given by
\begin{eqnarray}
|\Phi_t\rangle &=& \alpha
|0\rangle_{\mathrm{c}_1}|\bar{\mathbf{0}}\rangle_{\mathrm{r}_1}
|0\rangle_{\mathrm{c}_2}|\bar{\mathbf{0}}\rangle_{\mathrm{r}_2} \nonumber \\
&&+\beta|\phi_t\rangle_{\mathrm{c}_1 \mathrm{r}_1}
|\phi_t\rangle_{\mathrm{c}_2 \mathrm{r}_2}.
\end{eqnarray}
We observe that the overall state evolves as a four-qubit system.
By tracing out the reservoir states, the reduced two-cavity reduced density matrix reads
\begin{equation}\label{rhoc1c2}
\rho_{\mathrm{c}_1 \mathrm{c}_2}=\left(
\begin{array}{cccc}
  \alpha^2+\beta^2\chi^4 & 0 & 0 & \alpha\beta\xi^2 \\
  0 & \beta^2\xi^2\chi^2 & 0 & 0 \\
  0 & 0 & \beta^2\xi^2\chi^2 & 0 \\
  \alpha\beta\xi^2 & 0 & 0 & \beta^2\xi^4 \\
\end{array}
\right).
\end{equation}
This reduced state $\rho_{\mathrm{c}_1 \mathrm{c}_2}$ has the
structure of an $X$  matrix and exhibits ESD for $\alpha <
\beta$~\cite{Yu04,Marcelo06}. On the other hand, when tracing out
cavity modes we are led to the reduced reservoir state
\begin{equation}\label{rhor1r2}
\rho_{\mathrm{r}_1 \mathrm{r}_2}=\left(
\begin{array}{cccc}
  \alpha^2+\beta^2\xi^4 & 0 & 0 & \alpha\beta\chi^2 \\
  0 & \beta^2\chi^2\xi^2 & 0 & 0 \\
  0 & 0 & \beta^2\xi^2\chi^2 & 0 \\
  \alpha\beta\chi^2 & 0 & 0 & \beta^2\chi^4 \\
\end{array}
\right),
\end{equation}
whose structure also corresponds to an $X$ state. When replacing
$\xi(t)\leftrightarrow \chi(t)$, this state is
complementary to the state in Eq.~(\ref{rhoc1c2}). If $\rho_{\mathrm{c}_1 \mathrm{c}_2}$ is exhibiting
ESD, what happens then with  $\rho_{\mathrm{r}_1 \mathrm{r}_2}$?  To
answer this question we calculate the
\emph{concurrence}~\cite{Wootters98} for $\rho_{\mathrm{c}_1
\mathrm{c}_2}$, which for the particular state is given by the
simple expression
\begin{equation}
\mathcal{C}(t)=\max\{0,-2\lambda\},
\end{equation}
with $\lambda$ being the negative eigenvalue of the density matrix
partial transpose. For reduced states $\rho_{\mathrm{c}_1
\mathrm{c}_2}$ and $\rho_{\mathrm{r}_1 \mathrm{r}_2}$ these negative
eigenvalues are given by
\begin{equation}
\lambda_{\mathrm{c}_1 \mathrm{c}_2}=e^{-\kappa
t}\big[\beta^2(1-e^{-\kappa t})-|\alpha \beta|\big], \label{lcc}
\end{equation}
\begin{equation}
\lambda_{\mathrm{r}_1 \mathrm{r}_2}=(1-e^{-\kappa
t})\big[\beta^2e^{-\kappa t} -|\alpha \beta|\big]. \label{lrr}
\end{equation}
Figure~\ref{figqubit} shows the evolution of concurrence between
the two cavities (solid line) and the two reservoirs (dashed
line). Despite the entanglement between the two cavities suddenly
disappears,  sudden birth of entanglement arises between the two
reservoirs. Note that the
entanglement contained initially in the cavity-cavity subsystem is
transferred to the bipartite reservoir system. The time for
which ESD and the entanglement sudden birth (ESB) occur can be
calculated from Eqs.~(\ref{lcc}) and (\ref{lrr}), looking for the
time where $\lambda_{\mathrm{c}_1 \mathrm{c}_2}$ becomes positive
for ESD and the time for which $\lambda_{\mathrm{r}_1
\mathrm{r}_2}$ becomes negative for ESB,

\begin{equation}
\begin{tabular}{l}
  $t_{\mathrm{ESD}} = -\frac{1}{\kappa} \ln{\left(1-\frac{\alpha}{\beta}\right)}$, \\
  \\
  $t_{\mathrm{ESB}} = \frac{1}{\kappa} \ln{\frac{\beta}{\alpha}}$.\label{timesqubits}
\end{tabular}
\end{equation}
From these expressions we learn that ESB occurs for $\beta >
\alpha$, as is the case for ESD. In other words, the presence of ESD
implies necessarily the apparition of ESB and, consequently,
asymptotic decay of entanglement between cavities implies an
asymptotic birth and growing of entanglement between reservoirs.

For the situation in Fig.~\ref{figqubit} we have
$t_{\mathrm{ESB}} < t_{\mathrm{ESD}}$. However, as can be easily
seen from Eqs.~(\ref{timesqubits}), when $\beta=2\alpha$,
$t_{\mathrm{ESB}} = t_{\mathrm{ESD}}$, that is, ESB and ESD happen simultaneously.
Furthermore, when  $\beta > 2\alpha$,  ESB occurs after ESD.
Although this is clear from Eqs.~(\ref{timesqubits}), it is not
necessarily intuitive. In fact, this condition yields a time
window where neither the cavity-cavity nor the
reservoir-reservoir subsystems have entanglement.

\begin{figure}[t]
\includegraphics[width=55mm]{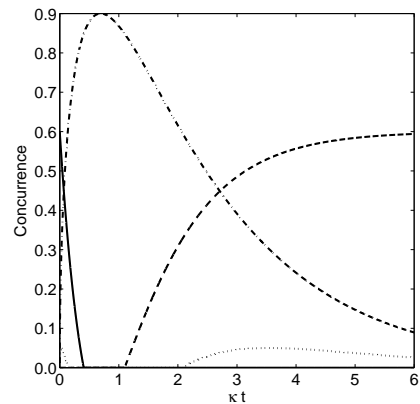}
\caption{Evolution of two-qubit concurrence for different
partitions: $\mathcal{C}_{\mathrm{c}_1 \mathrm{c}_2}$ (solid
line), $\mathcal{C}_{\mathrm{r}_1 \mathrm{r}_2}$ (dashed line),
$\mathcal{C}_{\mathrm{c}_1 \mathrm{r}_1}$ (dot-dashed line),
$\mathcal{C}_{\mathrm{c}_1 \mathrm{r}_2}$ (dotted line), for the
initial state of Eq.~(\ref{cavent}) with $\alpha=1/\sqrt{10}$ and
$\beta=3/\sqrt{10}$.} \label{figqubitpart}
\end{figure}

To have an idea of how the entanglement is shared among the parties, we
study  the entanglement present in different partitions. We start considering all bipartite partitions of two qubits,
namely: $c_1\otimes c_2$, $r_1\otimes r_2$, $c_1\otimes r_1$ and
$c_1\otimes r_2$, as shown in Fig.~\ref{figqubitpart}.
\begin{figure}[t]
\hspace*{-1cm}\includegraphics[width=70mm]{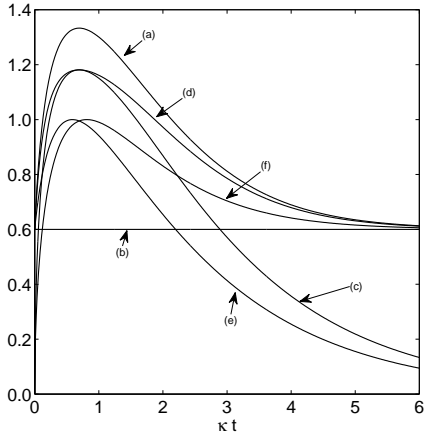}
\caption{Evolution of entanglement for different partitions: (a)
$c_1\otimes r_1 \otimes c_2 \otimes r_2 $; (b) $(c_1\otimes r_1)
\otimes (c_2 \otimes r_2) $; (c) $(c_1\otimes c_2) \otimes (r_1
\otimes r_2)$; (d) $(c_1\otimes r_2) \otimes (c_2 \otimes r_1) $;
(e) $c_1\otimes (r_1 \otimes c_2 \otimes r_2) $; (f) $r_1\otimes
(c_1 \otimes c_2 \otimes r_2) $ for the initial state of
Eq.~(\ref{cavent}) with $\alpha=1/\sqrt{10}$ and
$\beta=3/\sqrt{10}$.} \label{figqubitpart2}
\end{figure}
In particular for partition $c_1\otimes
r_1$, the entanglement is given by
\begin{equation}\label{c1r1}
\mathcal{C}_{\mathrm{c}_1 \mathrm{r}_1 }(t)=2\beta^2
\sqrt{(1-e^{-\kappa t})e^{-\kappa t}}.
\end{equation}
In the region where there is no entanglement, that is,
$\mathcal{C}_{\mathrm{c}_1 \mathrm{c}_2}=\mathcal{C}_{\mathrm{r}_1
\mathrm{r}_2}=0$,  entanglement between a cavity and its
corresponding reservoir $\mathcal{C}_{\mathrm{c}_1
\mathrm{r}_1 }(t)$ reaches its maximum value. This fact is
independent of the initial probability amplitudes $\alpha$ and
$\beta$ and occurs for a time $t=\kappa^{-1}\ln{(2)}$ which
corresponds also to the time when
$t_{\mathrm{ESD}}=t_{\mathrm{ESB}}$.

Entanglement of other bipartite partitions is shown in
Fig.~\ref{figqubitpart2}(b)-(f), and the multipartite
entanglement between the four effective qubits in
Fig.~\ref{figqubitpart2}(a). Such entanglement is described
by the multipartite concurrence
$\mathcal{C}_N$~\cite{Carvalho04}. For partitions (b)-(f) the
entanglement is obtained through the square root of the
\emph{tangle}~\cite{Rungta03} which in the pure two-qubit case
coincides with the concurrence. Note that $\mathcal{C}_N$ has the
same value at $t=0$ and $t\rightarrow\infty$, showing complete
entanglement transfer from cavities to reservoirs.

Although entanglement transfer from cavities to reservoirs is
mediated only by the interaction of each cavity and its
corresponding reservoir, entanglement may also flow
through other parties. Figure~\ref{figqubitpart2}(b) shows that the partition
$(c_1\otimes r_1) \otimes (c_2 \otimes r_2) $ has constant
entanglement. However,
Fig.~\ref{figqubitpart} shows that entanglement in the
two-qubit partition $c_1\otimes r_2$ is created along the
evolution, implying that entanglement flows also to the noninteracting
partitions. This fact can be visualized as follows: initially the entanglement is contained in the partition $c_{1} \otimes c_{2}$. Then, due to the interaction between cavities and reservoirs, for example $c_{1}$ and $r_{1}$, the information about the quantum state of $c_{1}$ begins to be mapped into the quantum state of $r_{1}$. Therefore, some of the quantum information contained in the joint-system of the cavities~\cite{Yu04} is now present in the joint-system of $c_{2}\otimes r_{1}$, producing entanglement in this partition.

It is interesting to investigate whether the features we
have analyzed so far are present for higher dimensional systems. For example, we consider the case of qutrit cavity states. Following similar steps used to obtain
Eq.~(\ref{phit1}), it is not difficult to calculate the evolution of
a single cavity mode, initially in a two-photon
$|2\rangle_\mathrm{c}$ state, interacting with the reservoir
initially in the vacuum state. The initial state
$|\phi_0^{(2)}\rangle=|2\rangle\otimes |\mathbf{\bar{0}}\rangle$
evolves according with
\begin{equation}\label{phit2}
|\phi _{t}^{(2)}\rangle =\xi^2 (t)|2\rangle
_{\mathrm{c}}|\bar{\mathbf{0}}\rangle _{\mathrm{r}}+\sqrt 2 \xi
(t)\chi (t)|1\rangle _{\mathrm{c}}|\mathbf{\bar{1}}\rangle
_{\mathrm{r}}+\vartheta (t)|0\rangle
_{\mathrm{c}}|\mathbf{\bar{2}}\rangle _{\mathrm{r}},
\end{equation}
where,
\begin{eqnarray}
|\mathbf{\bar{2}}\rangle _{\mathrm{r}}&=&\frac{1}{\vartheta
(t)}\bigg(\sum_{k=1}^{N} |\lambda _{k}(t)|^2 |2_{k}\rangle\\
\notag &&+\sqrt 2 \sum_{k\neq q =1}^{N} \lambda _{k}(t)\lambda
_{q}(t) |1_{k}\dots 1_{q}\rangle _{\mathrm{r}}\bigg),
\end{eqnarray}
and $\vartheta(t)=\sqrt{1-\xi^4(t)-2\xi^2(t)\chi^2(t)}$. We can now
study the entanglement when the initial state is given by
\begin{equation}\label{qutritsini}
|\Phi_0\rangle=\left(\alpha|0\rangle_{\mathrm{c}_1}
|0\rangle_{\mathrm{c}_2}+\beta|1\rangle_{\mathrm{c}_1}
|1\rangle_{\mathrm{c}_2}+\gamma|2\rangle_{\mathrm{c}_1}
|2\rangle_{\mathrm{c}_2}\right)\otimes
|\bar{\mathbf{0}}\rangle_{\mathrm{r}_1}|\bar{\mathbf{0}}\rangle_{\mathrm{r}_2}.
\end{equation}

As no entanglement monotone exists for an arbitrary higher
dimensional state, we focus on the analytical expression for a
lower bound of entanglement (LBOE) found by Chen, \emph{et.
al.}~\cite{albeverio05}, based on the PPT~\cite{Peres,Horodecki}
and realignment criterion~\cite{Rudolph,Chen}. The LBOE monotone
of a bipartite system (A and B) denoted $\Lambda$ is given by
$\Lambda =\max \left(\left\Vert \rho ^{T_{A}}\right\Vert
,\left\Vert R(\rho )\right\Vert \right)$, where the trace norm
$\left\Vert \cdot \right\Vert$ is defined as $\left\Vert
G\right\Vert =tr(GG^{\dagger })^{\frac{1}{2}}$. The matrix $\rho
^{T_{A}}$ is the partial transpose with respect to the subsystem
$A$, that is, $\rho _{ik,jl}^{T_{A}}=\rho _{jk,il}$, and the
matrix $R(\rho )$ is realignment matrix defined as $R(\rho
)_{ij,kl}=\rho _{ik,jl}$. The values of $\Lambda$ ranges from 1
(separable state) to $d$ (maximally entangled), where $d$ is the
dimension of the lower dimensional subsystem.

\begin{figure}[t]
\includegraphics[width=55mm]{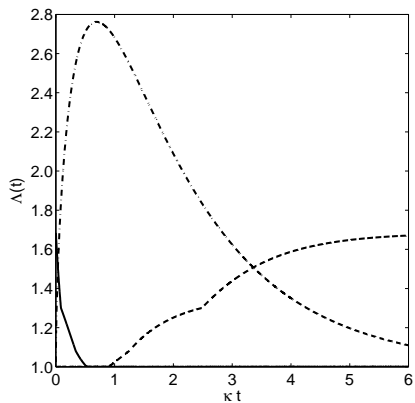}
\caption{Evolution of two-qutrit LBOE for different partitions:
$\Lambda_{\mathrm{c}_1 \mathrm{c}_2}$ (solid line),
$\Lambda_{\mathrm{r}_1 \mathrm{r}_2}$ (dashed line),
$\Lambda_{\mathrm{c}_1 \mathrm{r}_1}$ (dot-dashed line),
$\Lambda_{\mathrm{c}_1 \mathrm{r}_2}$ (dotted line), for the
initial state of Eq.~(\ref{qutritsini}) with $\alpha=1/\sqrt{38}$,
and $\beta=1/\sqrt{38}$ and $\gamma=6/\sqrt{38}$.}
\label{figqutrit}
\end{figure}
In Fig.~(\ref{figqutrit}), the evolution of $\Lambda_{\mathrm{c}_1
\mathrm{c}_2 }(t)$ and $\Lambda_{\mathrm{r}_1 \mathrm{r}_2}(t)$ is
shown. We observe that the sudden death of the cavity-cavity
entanglement is accompanied by  sudden birth of
reservoir-reservoir entanglement as in the two-qubit case.
Moreover, the LBOE dynamics between the reservoirs exhibits abrupt
changes as the LBOE between cavities~\cite{lastra07}. The times
for the ESD and the ESB to appear are
\begin{eqnarray}\label{timesqutrits}
  t_{\mathrm{ESD}}^{\mathrm{c}_1 \mathrm{c}_2} &=& -\frac{1}{\kappa} \ln{\left(1-\left(\frac{\alpha}{\gamma}\right)^{\frac{1}{2}}\right)} , \\
  t_{\mathrm{ESB}}^{\mathrm{r}_1 \mathrm{r}_2} &=& \frac{1}{2 \kappa} \ln{\frac{\gamma}{\alpha}} .
\end{eqnarray}
As for the two-qubit case, the time for wich ESD and ESB occur
simultaneously results to be $t=\kappa^{-1} \ln{(2)}$.
\begin{figure}[t]
\includegraphics[width=55mm]{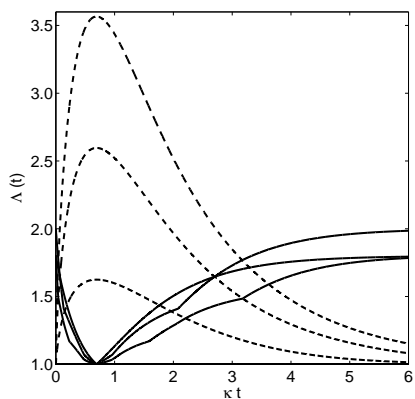}
\caption{Evolution of the LBOE $\Lambda_{\mathrm{c}_1
\mathrm{c}_2}$ and $\Lambda_{\mathrm{r}_1 \mathrm{r}_2}$ (solid
lines) and $\Lambda_{\mathrm{c}_1 \mathrm{r}_1}$ (dashed lines)
for initial state of Eq.~(\ref{ddim}) with $d=2,3$, and $4$.
Probability amplitudes $\alpha_k$ with $k=0,1,..,d-1$ are all
equals to $1/\sqrt{\sum_{k=0}^{d} |\alpha_k|^2}$ and
$\alpha_d=2^{d-1}\alpha_0$.} \label{tconst}
\end{figure}

In general, for a $d \otimes d$-dimensional bipartite system, each
one coupled to an independent reservoir, and initially prepared in
a state of the form
\begin{equation}\label{ddim}
|\Psi_0\rangle=\sum_{k=0}^d \alpha_k
|k\rangle_{\mathrm{c}_1}|k\rangle_{\mathrm{c}_2} \otimes
|\bar{\mathbf{0}}\rangle_{\mathrm{r}_1}
|\bar{\mathbf{0}}\rangle_{\mathrm{r}_2},
\end{equation}
we have numerically observed that the time when
$t_{\mathrm{ESD}}=t_{\mathrm{ESB}}$ does not depend on the dimension
of the systems. As can be seen from Fig. \ref{tconst} the time for
which  $t_{\mathrm{ESD}}=t_{\mathrm{ESD}}=\kappa^{-1} \ln{(2)}$.
The necessary condition for these times to be equal is
$\alpha_d/\alpha_0=2^{d-1}$. Although this condition does not
depend on the remaining probability amplitudes $\alpha_k$ with $k
\neq 0,d$, the condition $\alpha_k < \alpha_d$ must be satisfied
to ensure the presence of ESD and ESB.

In conclusion, we have shown that ESD in a bipartite system independently coupled to two
reservoirs is necessarily related to the ESB between the environments. The loss of entanglement is related to the birth of entanglement between the reservoirs and other partitions. We analytically demonstrate that ESD and ESB occur at times depending
on the amplitudes of the initial entangled state. We found that ESB occur before, together, or even after ESD. In the latter case,
when neither cavities nor reservoirs have entanglement, we have analyzed how the entanglement flows to other partitions. Finally, we showed that the simultaneous occurrence of ESD and ESB is independent of the system dimension.

C.E.L. acknowledges financial support from Fondecyt 11070244, DICYT USACH and PBCT-CONICYT PSD54, F.L. from Fondecyt 3085030, G.R. from CONICYT grants, J.C.R. from Fondecyt 1070157 and Milenio ICM P06-067, E.S. from SFB~631, EU EuroSQIP, and Ikerbasque Foundation.

\end{document}